\documentclass[aps,prl,superscriptaddress,showpacs,twocolumn]{revtex4}
\usepackage{amsmath,amsfonts,bm,graphicx,amssymb}
\newcommand{\W}{\mathcal{W}}
\newcommand{\Wbar}{\mathcal{\overline{W}}}

\newcommand{\QQ}{\mathcal{Q}}
\newcommand{\RR}{\mathcal{R}}
\newcommand{\hatrho}{\hat{\rho}}
\newcommand{\hatgamma}{\hat{\gamma}}
\newcommand{\hatH}{\hat{H}}
\newcommand{\hatsigma}{\hat{\sigma}}
\newcommand{\llangle}{\langle\!\langle}
\newcommand{\rrangle}{\rangle\!\rangle}
\newcommand{\rlangle}{\rangle\!\langle}
\newcommand{\rrllangle}{\rrangle\!\llangle}
\begin{document}

\title{Counting Statistics of Non-Markovian Quantum Stochastic Processes}
\author{Christian Flindt}
\affiliation{Laboratory of Physics, Helsinki University of
Technology, P.\ O.\ Box 1100, 02015 HUT, Finland} \affiliation{Department of Micro- and Nanotechnology,
             Technical University of Denmark, DTU Nanotech, Building 345 East,
             2800 Kongens Lyngby, Denmark}
\author{Tom\'{a}\v{s} Novotn\'{y}}
\affiliation{Department of Condensed Matter Physics,
             Faculty of Mathematics and Physics, Charles University,
             Ke Karlovu 5, 12116 Prague, Czech Republic}
\author{Alessandro Braggio}
\affiliation{LAMIA-INFM-CNR,
             Dipartimento di Fisica,
             Universit\`{a} di Genova,
             Via Dodecaneso 33, 16146 Genova, Italy}
\author{Maura Sassetti}
\affiliation{LAMIA-INFM-CNR,
             Dipartimento di Fisica,
             Universit\`{a} di Genova,
             Via Dodecaneso 33, 16146 Genova, Italy}
\author{Antti-Pekka Jauho}
\affiliation{Laboratory of Physics, Helsinki University of
Technology, P.\ O.\ Box 1100, 02015 HUT, Finland} \affiliation{Department of Micro- and Nanotechnology,
             Technical University of Denmark, DTU Nanotech, Building 345 East,
             2800 Kongens Lyngby, Denmark}

\date{\today}
\begin{abstract}
We derive a general expression for the cumulant generating function
(CGF) of non-Markovian quantum stochastic transport processes.  The
long-time limit of the CGF is determined by a single dominating pole
of the resolvent of the memory kernel from which we extract the
zero-frequency cumulants of the current using a recursive scheme.
The finite-frequency noise is expressed not only in terms of the
resolvent, but also initial system-environment correlations. As an
illustrative example we consider electron transport through a
dissipative double quantum dot for which we study the effects of
dissipation on the zero-frequency cumulants of high orders and the
finite-frequency noise.
\end{abstract}



\maketitle

Full counting statistics (FCS) has recently attracted intensive
theoretical \cite{Nazarov:2003} and experimental \cite{Reulet:2003}
attention. The interest stems from the usefulness of FCS as a
sensitive diagnostic tool of stochastic electron transport through
mesoscopic systems. Detectable mechanisms include quantum-mechanical
coherence, entanglement, disorder, and dissipation
\cite{Nazarov:2003}. Mathematically, FCS encodes the complete
knowledge of the probability distribution $P(n,t)$ of the
transmitted number $n$ of electrons or, equivalently, of all
corresponding cumulants. Non-zero higher order cumulants describe
non-Gaussian behavior. The study of counting statistics for
stochastic processes in general is of broad relevance for a wide
class of problems, also outside mesoscopic physics. For example,
rare events, whose study has become an important topic within
non-equilibrium statistics of stochastic systems in physics,
chemistry, and biology \cite{Sevick:2007}, are reflected in higher
order cumulants. Efficient methods for evaluating  the counting
statistics of stochastic processes are therefore of urgent need.

In this Letter, we present a method which unifies and extends a
number of earlier approaches to FCS within a generalized master
equation (GME) formulation
\cite{Bagrets:2003,Flindt:2005,Braggio:2006}. The earlier approaches
have in practice been limited to systems with only a few states
\cite{Bagrets:2003,Braggio:2006}, or only the first few current
cumulants \cite{Flindt:2005}.  In contrast, our theory enables
studies of a much larger class of problems: Evaluation of
zero-frequency current cumulants of very high orders for
non-Markovian systems with many states is now possible. Furthermore,
the method allows us to develop a general approach to the
finite-frequency current noise of non-Markovian transport processes.
In the case of finite-frequency noise, we show that not only the
memory kernel but also initial system-environment correlations are
crucial. Such correlations can be, and have been
\cite{Braggio:2006,Aguado:2004a}, neglected for non-Markovian
processes at low frequencies, but must be included at frequencies
comparable with the internal frequencies of the system. We
demonstrate our methods on a system of recent experimental
relevance, namely transport through a dissipative double quantum dot
\cite{Aguado:2004a,Kiesslich:2007}, but they may easily be applied
to other electronic (or photonic) counting systems, as well as other
counted quantities, such as heat or work, in non-equilibrium systems
\cite{Seifert:2007}.

\emph{Non-Markovian GME.} Consider a nanoscale transport system
governed by a generic non-Markovian GME of the form
\cite{Zwanzig:2001,Makhlin:2001}
\begin{equation}
\frac{d}{dt}\hatrho(n,t)=\sum_{n'}\int_{0}^{t}dt'\W(n-n',t-t')\hatrho(n',t')+\hatgamma(n,t).
\label{eq:GME}
\end{equation}
Here, the reduced density matrix of the system $\hatrho(t)$ has been
resolved into components $\hatrho(n,t)$ corresponding to the number
of electrons $n$ passing through the nanosystem within time-span
$[0,t]$. The memory kernel $\W$ describes the influence of the
environment on the dynamics of the system, while the inhomogeneity
$\hatgamma$ accounts for initial correlations between system and
environment. Both $\W$ and $\hatgamma$ decay with time, usually on a
comparable timescale, so that $\hatgamma$ is irrelevant for the
long-time limit. The inhomogeneity $\hatgamma$ does, however, play a
crucial role at finite times. The probability distribution for the
number of transferred charges is
$P(n,t)=\mathrm{Tr}\{\hatrho(n,t)\}$. The corresponding cumulant
generating function (CGF) $\mathcal{S}(\chi,t)$ is defined as
$e^{\mathcal{S}(\chi,t)}\equiv\sum_{n}P(n,t)e^{in\chi}$. In Laplace
space Eq.\ (\ref{eq:GME}) leads to the algebraic expression
$\hatrho(\chi,z)=\mathcal{G}(\chi,z)[\hatrho(\chi,t=0)+\hatgamma(\chi,z)]$,
where $\mathcal{G}(\chi,z)\equiv[z-\W(\chi,z)]^{-1}$ is the
resolvent of the kernel, and
$\hatrho(\chi,z)\equiv\sum_{n}\int_0^{\infty}dt\hatrho(n,t)e^{in\chi-zt}$
and similarly for $\hatgamma(\chi,z)$ and $\W(\chi,z)$. Inverting
the Laplace transformation, the CGF then becomes
\begin{equation}
e^{\mathcal{S}(\chi,t)}=\frac{1}{2\pi
i}\int_{a-i\infty}^{a+i\infty}dz\langle\mathcal{G}(\chi,z)\rangle
e^{zt}, \label{eq:CGF}
\end{equation}
where $a$ is a real number, chosen such that all singularities of
the integrand are situated to the left of the vertical line of
integration. We have moreover introduced the notation
$\langle\mathcal{G}(\chi,z)\rangle\equiv\mathrm{Tr}\{\mathcal{G}(\chi,z)[\hatrho(\chi,t=0)+\hatgamma(\chi,z)]\}$.
Equation (\ref{eq:CGF}) contains the full statistical information
about the charge transfer process. It is a powerful formal result,
but it also leads to useful practical schemes, as we shall now
demonstrate.

\emph{Zero-frequency FCS.} Consider first the zero-frequency
cumulants of the current, defined as $\llangle
I^m\rrangle=\frac{d}{dt}\frac{\partial^m\mathcal{S}(\chi,t)}{\partial{(i\chi)^m}}|_{\chi\rightarrow
0, t\rightarrow\infty}$, $m=1,2,\ldots$. We assume that the system
with the counting field $\chi$ set to zero tends exponentially to a
unique stationary state determined by the $1/z$ pole of the
resolvent $\mathcal{G}(\chi=0,z)$. The stationary state is given by
the eigenvector corresponding to the zero-eigenvalue of
$\W\equiv\W(\chi=0,z=0)$, i.e., $\lim_{t\rightarrow
\infty}\hatrho(t)=\hatrho^{\rm stat}$, where $\hatrho^{\rm stat}$ is
the normalized solution to $\W\hatrho^{\rm stat}=0$. With finite
values of $\chi$, an eigenvalue $\lambda_0(\chi,z)$ develops
adiabatically from the zero-eigenvalue, such that
$\lambda_0(0,z)=0$, and the long-time behavior is determined by the
isolated pole  structure $1/[z-\lambda_0(\chi,z)]$ of
$\mathcal{G}(\chi,z)$ close to zero. This pole $z_0(\chi)$ solves
\begin{equation}
z_0-\lambda_0(\chi,z_0)=0, \label{eq:self-consistency}
\end{equation}
and goes to zero with $\chi$ going to zero, i.e., $z_0(0)=0$. We
thus find $e^{\mathcal{S}(\chi,t)}\rightarrow
\mathcal{D}(\chi)e^{z_0(\chi)t}$ for large $t$, where
$\mathcal{D}(\chi)$ is a time-independent function depending on the
initial conditions and correlations. The current cumulants then read $\llangle
I^m\rrangle=\frac{\partial^m
z_0(\chi)}{\partial{(i\chi)^m}}|_{\chi\rightarrow 0}$. In the
Markovian limit for the kernel  $\W(\chi,z\rightarrow 0)$ we get
$z_0(\chi)=\lambda_0(\chi,0)$ as found in Refs.~\cite{Bagrets:2003,
Flindt:2005}.

\emph{Recursive scheme.} When the involved matrices are large, it
may be non-trivial to determine the full $\chi$ and $z$ dependence
of the eigenvalue $\lambda_0(\chi,z)$, and thereafter solve Eq.\
(\ref{eq:self-consistency}). Instead, we expand the eigenvalue as
$\lambda_0(\chi,z)=\sum_{k,l=0}^{\infty}\frac{(i\chi)^k}{k!}\frac{z^l}{l!}c^{(k,l)}$
with $c^{(0,L)}=0$, and calculate the expansion coefficients
recursively using Rayleigh-Schr\"{o}dinger perturbation theory
\cite{Flindt:2008}:
\begin{equation}
\begin{split}
c^{(K,L)}=&\sum_{k=0}^K{K\choose k}\sum_{l=0}^L{L \choose
l}\llangle\tilde{0}|\Wbar^{(k,l)}|0^{(K-k,L-l)}\rrangle,\\
|0^{(K,L)}\rrangle = &\mathcal{R}\sum_{k=0}^K{K \choose k}\sum_{l=0}^L{L
\choose
l}[c^{(k,l)}-\Wbar^{(k,l)}]|0^{(K-k,L-l)}\rrangle,
\end{split}
\label{eq:recursivescheme}
\end{equation}
with $K,L=0,1,2,\ldots$. Here, $\llangle\tilde{0}|$ solves
$\llangle\tilde{0}|\W=0$, while $|0^{(0,0)}\rrangle$ is the
stationary state $\hatrho^{\rm stat}$. Moreover,
$\Wbar(\chi,z)\equiv\W(\chi,z)-\W$ has been expanded as
$\Wbar(\chi,z)=\sum_{k,l=0}^{\infty}\frac{(i\chi)^k}{k!}\frac{z^l}{l!}\Wbar^{(k,l)}$
with $\Wbar^{(0,0)}= 0$. Finally, the pseudoinverse of the kernel is
$\mathcal{R}\equiv\QQ\W^{-1}\QQ$ with $\QQ\equiv
1-|0^{(0,0)}\rrllangle\tilde{0}|$ \footnote{Details of the notation
can be found in Ref.\ \cite{Flindt:2004}.}. With the $c^{(K,L)}$'s
at hand we can solve Eq.\ (\ref{eq:self-consistency}) for
$z_0(\chi)$ to a given order in $\chi$, and from the expansion
$z_0(\chi)=\sum_{n=1}^{\infty}\frac{(i\chi)^n}{n!}\llangle
I^n\rrangle$ extract the zero-frequency cumulants of the current:
\begin{equation}
\begin{split}
\llangle
I^N\rrangle=&N!\sum_{k,l=0}^N\frac{1}{k!}\frac{1}{l!}P^{(N-k,l)}c^{(k,l)},\\
P^{(K,L)}=&\sum_{n=1}^K\frac{\llangle I^n\rrangle}{n!}P^{(K-n,L-1)}
\end{split}
\label{eq:recursivescheme2}
\end{equation}
with $L=0,1,2,\ldots$, and $K,N=1,2,3,\ldots$. For the auxiliary
quantity $P^{(K,L)}$, we have $P^{(K,0)}=\delta_{K,0}$,
$P^{(0,L)}=\delta_{0,L}$, and $P^{(K,-1)}\equiv 0$.

We illuminate the recursive scheme by evaluating the first three
cumulants of the current using Eq.\ (\ref{eq:recursivescheme2}); the
mean current, the variance (the noise), and the skewness:
\begin{equation}
\begin{split}
\llangle
I^1\rrangle =& c^{(1,0)},\\
\llangle I^2\rrangle =& c^{(2,0)}+2c^{(1,0)}c^{(1,1)},\\
\llangle I^3\rrangle =& c^{(3,0)}+3 c^{(2,0)}c^{(1,1)}\\
&+3c^{(1, 0)}\left[c^{(1, 0)} c^{(1, 2)}+2(c^{(1,1)})^2+ c^{(2,1)}\right].\\
\end{split}
\end{equation}
Higher order cumulants can be obtained in a similar manner,
analytically or numerically. Coefficients of the form $c^{(L,0)}$
are purely Markovian quantities, and the mean current is thus not
sensitive to non-Markovian effects, whereas higher order cumulants
are \cite{Braggio:2006,Engel:2004}. From Eq.\
(\ref{eq:recursivescheme}) we find for the coefficients $c^{(K,L)}$,
e.g., $c^{(1,0)}=\llangle\tilde{0}|\Wbar^{(1,0)}|0^{(0,0)}\rrangle$,
$c^{(1,1)}=\llangle\tilde{0}|(\Wbar^{(1,1)}-\Wbar^{(1,0)}\RR\Wbar^{(0,1)})|0^{(0,0)}\rrangle$,
and
$c^{(2,0)}=\llangle\tilde{0}|(\Wbar^{(2,0)}-2\Wbar^{(1,0)}\RR\Wbar^{(1,0)})|0^{(0,0)}\rrangle$.
Evaluation of the pseudoinverse $\RR$ amounts to solving matrix
equations which is feasible even with very large matrices
\cite{Flindt:2004}. Numerically, the recursive scheme is stable for
very high orders of cumulants ($>20$) as we have tested on simple
models.

\emph{Double quantum dot.} We illustrate our method by considering a
model of charge transport through a Coulomb blockaded double quantum
dot in a dissipative environment \cite{Aguado:2004a}. Maximally one
additional electron is allowed on the double quantum dot. The
Hamiltonian of the double quantum dot is
$\hatH_S=\frac{\varepsilon}{2}\hat{s}_z+T_c\hat{s}_x$, where the
pseudo-spin operators are $\hat{s}_z\equiv |L\rlangle L|-|R\rlangle
R|$ and $\hat{s}_x\equiv |L\rlangle R|+|R\rlangle L|$, respectively.
The tunnel coupling between the two quantums dot levels $|L\rangle$
and $|R\rangle$ is $T_c$, while $\varepsilon$ is their detuning. The
pseudo-spin system is tunnel-coupled to left ($L$) and right ($R$)
leads via the tunnel-Hamiltonian
$\hatH_T=\sum_{k_{\alpha},\alpha=L,R}(V_{k_{\alpha}}\hat{c}^{\dagger}_{k_\alpha}|0\rlangle
\alpha|+\mathrm{h.c.})$, with both leads described as
non-interacting fermions, i.e.,
$\hatH_{\alpha}=\sum_{k_\alpha}\varepsilon_{k_\alpha}\hat{c}^{\dagger}_{k_\alpha}\hat{c}_{k_\alpha},
\alpha=L,R $. Dissipation is provided by a reservoir of
non-interacting bosons that couple to the $\hat{s}_z$ component of
the pseudo-spin. The Hamiltonian is then
$\hatH=\hatH_S+\hatH_T+\hatH_L+\hatH_R+\hatH_B+\hat{V}_B\hat{s}_z$,
where $\hatH_B=\sum_j\hbar\omega_j\hat{a}_j^{\dagger}\hat{a}_j$ and
$\hat{V}_B= \sum_j \frac{g_j}{2}(\hat{a}_j^{\dagger}+\hat{a}_{j})$.

To describe charge transport through the system we trace out the
leads following Gurvitz and Prager \cite{Gurvitz:1996}, leading to
an equation of motion for the reduced density matrix
$\hatsigma=(\hatsigma_{00},\hatsigma_{LL},\hatsigma_{RR},\hatsigma_{LR},\hatsigma_{RL})^T$
of the double dot and the bath of bosons. The elements
$\hatsigma_{ij}$  are still operators in the Hilbert space of the
boson bath. Charges are assumed to enter the left quantum dot from
the left lead and leave from the right quantum dot via the right
lead with energy-independent rates
$\Gamma_{\alpha}(\epsilon)=2\pi\sum_{k}|V_{k_{\alpha}}|^2\delta(\epsilon-\varepsilon_{k_\alpha})=\Gamma_{\alpha}$,
$\alpha=L,R$. This approach is valid to all orders in the tunnel
coupling $T_c$ under the assumption of a large bias across the
system \cite{Gurvitz:1996}.

Next, we consider the electronic occupation probabilities
$\rho_{i}=\mathrm{Tr}_B\{\hatsigma_{ii}\}$, $i=0,L,R$, where
$\mathrm{Tr}_B$ is a trace over the bosonic degrees of freedom. A
closed system of equations is obtained by assuming that the boson
bath at any time is in local equilibrium corresponding to the given
charge state: $\hatsigma_{ii}\simeq
\rho_{i}\otimes\hatsigma_i(\beta)$, $i=L,R$, where
$\hatsigma_{L/R}(\beta)\equiv e^{-\beta
H_B^{(\pm)}}/\mathrm{Tr}_B\{e^{-\beta H_B^{(\pm)}}\}$,
$H_B^{(\pm)}\equiv H_B\pm V_B$, and $\beta=1/k_BT$ is the inverse
temperature (see e.g.\ Sec.\ IV C in Ref.\ \cite{Flindt:2004}). This
approximation is valid when the bath-assisted hopping rates
$\Gamma_B^{(\pm)}(z)$ (proportional to $T_c^2$, see below) are much
smaller than $\Gamma_{L/R}$. The memory kernel for this model, with
$\hatrho=(\rho_{0},\rho_{L},\rho_{R})^T$, then reads
\begin{equation}
\W(\chi,z)=
\begin{pmatrix}
  -\Gamma_L & 0     & \Gamma_Re^{i\chi}\\
  \Gamma_L  & -\Gamma_{B}^{(+)}(z)     & \Gamma_{B}^{(-)}(z)         \\
  0         & \Gamma_{B}^{(+)}(z)     & -\Gamma_{B}^{(-)}(z)-\Gamma_R \\
\end{pmatrix}.
\label{eq:kernel}
\end{equation}
Here, the counting field $\chi$ has been introduced in the
off-diagonal element containing the rate
$\Gamma_R\rightarrow\Gamma_Re^{i\chi}$, corresponding to counting of
the number of electrons that have been collected in the right lead.
The bath-assisted hopping rates entering the kernel are
$\Gamma_{B}^{(\pm)}(z)=T_c^2[g^{(+)}(z_\pm)+g^{(-)}(z_\mp )]$ with
$g^{(\pm)}(z)=\int_0^{\infty}dt e^{-W(\mp t)-zt}$, $W(t)=\int_0^{\infty}d\omega
J(\omega)\{[1-\cos(\omega t)]\coth{(\beta\omega/2)}+i\sin(\omega
t)\}/\omega^2$, and $z_\pm=z\pm i\varepsilon+\Gamma_R/2$. The
spectral function of the heat bath is
$J(\omega)\equiv\sum_j|g_j|^2\delta(\omega-\omega_j)$, and below we
show results for Ohmic dissipation, $J_\Omega(\omega)=2\alpha \omega
e^{-\omega/\omega_c}$, when the rates can be evaluated either
analytically (for $\beta\omega_c\gg 1$) or numerically.

\begin{figure}
\begin{center}
\includegraphics[width=0.48\textwidth]{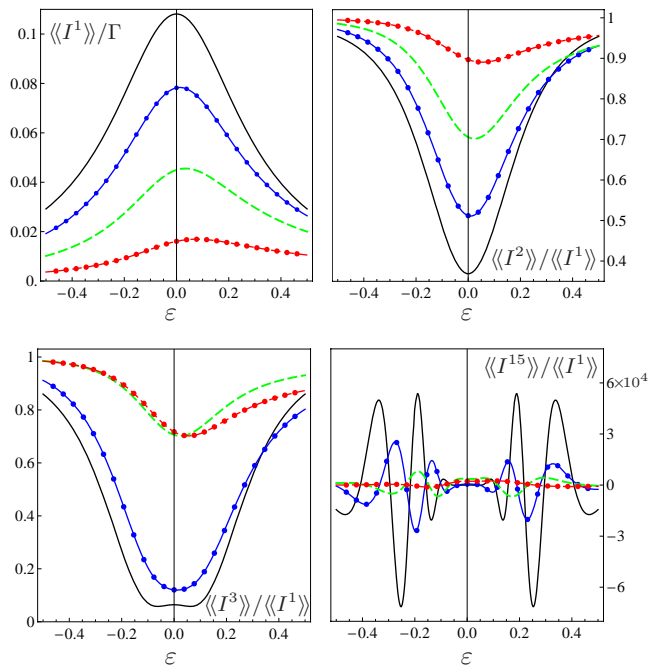}
\caption{(color online). Zero-frequency cumulants of the current as
functions of the level detuning $\varepsilon$ with different
dissipation strengths $\alpha$. Parameters are
$\Gamma=\Gamma_L=\Gamma_R=0.5$, $T_c=0.1$, $k_BT=0$, $\omega_c=500$,
and $2\pi\alpha=0$ (full line), 0.2 (full line with dots), 0.5
(dashed line), 1 (dashed line with dots). A large bias is applied
across the system.} \label{fig:ZeroFreqFig}
\end{center}
\end{figure}

In Fig.\ \ref{fig:ZeroFreqFig} we show the first three cumulants of
the current as functions of the level detuning $\varepsilon$ with
different dissipations strengths $\alpha$. As an illustrative
example the 15'th cumulant of the current, $\llangle
I^{15}\rrangle$, is also shown. As the dissipation strength is
increased, a clear suppression of the coherent features (with
$\alpha=0$) is seen. The increased coupling to the heat bath tends
to localize the electron to one of the two quantum dots, thereby
suppressing the effects of the coherent coupling between them. As a
result a cross-over from coherent to sequential tunneling is
observed with increasing $\alpha$. For large $\alpha$'s, the
sequential tunneling process between the two quantum dots
constitutes a `bottle-neck' and the cumulants consequently approach
the Poisson limit $\llangle I^m\rrangle/\llangle I^1\rrangle=1$,
$m=1,2,3,\ldots$. The typical behavior of cumulants is, however,
factorial growth, i.e., $|\llangle I^m\rrangle|\approx c\, q^m\, m!$
for some constants $c,\, q>0$. In the coherent case ($\alpha=0$),
this behavior is clearly seen for the 15'th cumulant, demonstrating
its high sensitivity to dephasing and decoherence mechanisms.

\emph{Finite-frequency noise.} The expression for the CGF, Eq.\
(\ref{eq:CGF}), allows us also to study the finite-frequency
spectrum of the second cumulant of the current, the (symmetrized)
current noise \footnote{Non-symmetrized noise in a non-Markovian
system was studied in Ref.\ \cite{Engel:2004}, but not within the
framework of FCS.}, expressed by MacDonald's formula as
$S_{II}(\omega)=\omega\int_{0}^{\infty}dt\sin{(\omega t)}\llangle
I^2\rrangle(t)$ \cite{MacDonald:1948,Flindt:2005b,Lambert:2007},
where $\llangle
I^2\rrangle(t)=\frac{d}{dt}\frac{\partial^2\mathcal{S}(\chi,t)}{\partial{(i\chi)^2}}|_{\chi\rightarrow
0}$. We then find
\begin{equation}
S_{II}(\omega)=-\frac{\omega^2}{2}\frac{\partial^2}{\partial{(i\chi)^2}}\left[\langle\mathcal{G}(\chi,z=
i\omega)\rangle+(\omega\rightarrow-\omega)\right]|_{\chi\rightarrow
0}.
\end{equation}
In order to evaluate this expression, we need to choose
$\hatrho(\chi,t=0)$ appropriately and find the inhomogeneity
$\hatgamma(\chi,z)$ as they enter the definition of
$\langle\mathcal{G}(\chi,z)\rangle$. Following Ruskov and Korotkov
\cite{Ruskov:2003} we assume that the system evolves from
$t_0=-\infty$, such that the
 electronic occupation probabilities at $t=0$, where electron counting begins, have reached the
stationary state, i.e., $\hatrho(n,t=0)=\delta_{n,0}\hatrho^{\rm
stat}$. For this model the inhomogeneity is independent of the
counting field \cite{Flindt:2008}:
\begin{equation}
\hatgamma(z)=\frac{\W-\W(\chi=0,z)}{z}\hatrho^{\rm stat}.
\end{equation}
We see that the effects of the initial correlations accounted for by
$\hatgamma(z)$ vanish in the long-time limit. Moreover, since
$\W\hatrho^{\rm stat}=0$, we find
$\langle\mathcal{G}(\chi,z)\rangle=\mathrm{Tr}\{\mathcal{G}(\chi,z)\mathcal{G}^{-1}(\chi=0,z)\hatrho^{\rm
stat}\}/z$ from which we can calculate the finite-frequency current
noise. We note that only the proper inclusion of the inhomogeneity
ensures a correct finite-time behavior, such as proper normalization
of $\mathrm{Tr}\{\hatrho(t)\}=\langle\mathcal{G}(\chi=0,t)\rangle=\mathrm{Tr}\{\hatrho^{\rm
stat}\}=1$ at all times.

\begin{figure}
\begin{center}
\includegraphics[width=0.49\textwidth]{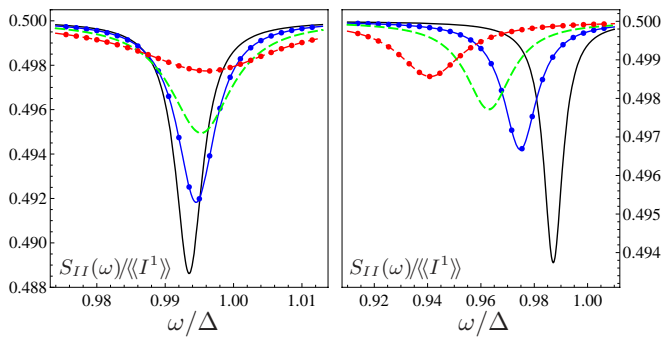}
\caption{(color online). Finite-frequency current noise spectrum for
different temperatures (left figure) and dissipation strengths
(right figure). Left figure: $\alpha=0.005$, $k_BT=0$ (full line), 1
(full line with dots), 2 (dashed line), 5 (dashed line with dots).
Right figure: $k_BT=0$, and $\alpha=0.01$ (full line), $0.02$ (full
line with dots), $0.03$ (dashed line), $0.05$ (dashed line with
dots). Other parameters are $\Gamma_L=\Gamma_R=0.01$, $T_c=3$,
$\varepsilon=10$, $\Delta=\sqrt{\varepsilon^2+(2T_c)^2}$, and
$\omega_c=500$. A large bias is applied across the system.}
\label{fig:FiniteFreqFig}
\end{center}
\end{figure}

Displacement currents (due to finite capacitances between quantum
dots and leads) can be included in the finite-frequency current
noise via the Ramo-Shockley theorem \cite{Blanter:2001}. Evaluating
the full current noise spectrum then requires an additional counting
field accounting for tunneling from the left lead
\cite{Lambert:2007}. For the shown results we have included
displacement currents and assumed identical capacitances between the
left/right quantum dot and the left/right lead. In Fig.\
\ref{fig:FiniteFreqFig} we show the finite-frequency current noise
$S_{II}(\omega)$ at frequencies around the hybridization energy
$\Delta=\sqrt{\varepsilon^2+(2T_c)^2}$, where signatures are
expected in the current noise spectrum
\cite{Aguado:2004a,Flindt:2005b,Blanter:2001}. Resonances are not
observed exactly at the `bare' value $\omega=\Delta$, but are
shifted towards lower frequencies. This renormalization occurs due
to coupling to the heat bath which dresses the eigenspectrum of the
electronic degrees of freedom, similar to the Lamb shift in atomic
physics. The left figure shows how the signatures due to the
coherent coupling between the two quantum dots are washed out with
increasing temperature. In the right figure, we show how the
frequency shift increases with increasing dissipation strength,
which simultaneously reduces the effect of the coherent coupling.
For larger $\alpha$'s ($\sim 0.1$) a change of line shape is
observed (not shown).

In conclusion, we have presented a general theory for current
fluctuations in non-Markovian quantum transport systems. Our methods
allow us to calculate recursively zero-frequency current cumulants
of very high orders, governed by a single dominating pole of the
resolvent of the memory kernel, as well as the finite-frequency
current noise, which is given not only by the resolvent, but also
initial correlations. As an illustrative example of our approach, we
have considered transport through a dissipative double quantum dot
for which we have studied the effects of dissipation and temperature
on the current cumulants of very high orders and the
finite-frequency current noise.

We would like to thank R.~Aguado, T.~Brandes, S.~Kohler, and
K.~Neto\v{c}n\'{y} for fruitful discussions and suggestions. The
work was supported by FiDiPro, Italian MIUR via PRIN05, ESF (`Arrays
of Quantum Dots and Josephson Junctions') and by the grant
202/07/J051 of the Czech Science Foundation.

\end{document}